\documentclass[conference]{IEEEtran}
\IEEEoverridecommandlockouts
\usepackage{cite}
\usepackage{amsmath,amssymb,amsfonts}
\usepackage{graphicx}
\usepackage{textcomp}
\usepackage{xcolor}
\usepackage{booktabs}
\usepackage{url}
\providecommand{\tightlist}{\setlength{\itemsep}{0pt}\setlength{\parskip}{0pt}}
\def\BibTeX{{\rm B\kern-.05em{\sc i\kern-.025em b}\kern-.08em
    T\kern-.1667em\lower.7ex\hbox{E}\kern-.125emX}}

\begin{document}

\title{Expecting Empathy: How Interaction Context Shapes Norms for Empathic Response in Digital Communication}

\author{
\IEEEauthorblockN{Tao Wang}
\IEEEauthorblockA{\textit{School of Information} \\
\textit{University of Toronto} \\
Toronto, Canada \\
taotw.wang@utoronto.ca}
\and
\IEEEauthorblockN{Chi-Ching Juan}
\IEEEauthorblockA{\textit{School of Information} \\
\textit{University of Toronto} \\
Toronto, Canada \\
sonia.juan@mail.utoronto.ca}
\thanks{Preprint. Under review.}
}

\maketitle

\begin{abstract}
A central challenge in affective computing is
determining appropriate empathy levels for different interaction
contexts. Prior work has characterized two poles: task-focused
interactions (empathy demand near zero) and emotional disclosure (high
empathy demand). This paper identifies a distinct intermediate type,
\emph{decision support under stress}, where a sender faces a
consequential choice while experiencing emotional difficulty. We
hypothesize that this type elicits an asymmetric empathy profile:
empathy comparable to emotional disclosure but instrumentality
comparable to task-focused exchange. We test five hypotheses using
28,239 post-reply dyads from three Reddit advice communities, classified
into three interaction types and scored for empathy depth, empathy form,
and instrumental proportion using LLM-based annotation with
pattern-based robustness checks. Results confirm the predicted
asymmetric profile: decision-support-under-stress replies show
significantly higher empathy than task-focused replies ($M = 0.47$
vs.\ $0.24$, $p < .001$) while maintaining high instrumentality
($0.83$ vs.\ $0.77$ for emotional disclosure, $p < .001$). Behavioral
empathy dominates (36.6\%), and community-validated response quality is
negatively associated with empathic expression ($r = -.075$, $p <
.001$). Community norms modulate baselines substantially but preserve
structural ordering. These findings establish a human empathy baseline
for this interaction type and have direct implications for calibrating
empathic expression in affective AI systems.
\end{abstract}


\section{INTRODUCTION}

The development of empathic AI systems has made significant progress in
recent years, with large language models (LLMs) now capable of
generating responses perceived as empathic \cite{b1, b2}. However, a
fundamental question has received comparatively less attention:
\emph{how much} empathy is appropriate across different interaction
contexts? Without empirically grounded benchmarks, AI systems default to
uniformly high empathy, an approach that can backfire, triggering
inauthenticity perceptions or emotional overreach \cite{b3, b4}.
Inaccurate empathy has been shown to be worse than minimal empathy,
underscoring the importance of calibration over maximization \cite{b5}.

Online advice communities provide a natural laboratory for studying
empathy calibration. Two poles are well characterized: task-focused
interactions where empathy is functionally irrelevant \cite{b6, b7},
and emotional disclosure where empathy is the primary currency \cite{b8, b9}. However, many real-world interactions fall between these poles.
When someone seeks career guidance while anxious, or requests financial
help while overwhelmed, they need both emotional recognition and
instrumental guidance.

We term this intermediate category \emph{decision support under stress}:
an interaction where (a) the sender faces a consequential choice, (b)
the sender is experiencing emotional difficulty related to the choice,
and (c) the message seeks both emotional recognition and instrumental
guidance. This type has been studied incidentally, as complicated cases
within clinical empathy \cite{b10} or edge cases within the emotional
support conversation (ESC) framework, but not identified as a distinct
type with its own empathy demand profile.

This paper makes three contributions. First, we identify decision
support under stress as a distinct interaction type with an empirically
observable asymmetric empathy profile. Second, we derive and test five
hypotheses about how human respondents calibrate empathy across
interaction types, community contexts, and response quality levels.
Third, we establish a large-scale human baseline (28,239 dyads) that can
inform the design of context-sensitive empathic AI systems.

\section{RELATED WORK}

\subsection{Empathy at the Two Poles}

Task-focused interaction has been studied since the development of
Interaction Process Analysis \cite{b11}. In online Q\&A platforms, norms
suppress socioemotional content \cite{b6}, and empathic messages are
largely absent in task-oriented communities \cite{b7}. The empathy demand
is near zero; introducing emotional content where none was offered
constitutes an epistemic error \cite{b12}.

At the emotional disclosure pole, felt understanding is central to
well-being \cite{b8}, perceived empathic effort predicts satisfaction more
strongly than actual accuracy \cite{b13}, and over-empathizing is
empirically detectable and consequential \cite{b3}. The demand profile is
high emotional richness, high contextual anchoring, and moderate
perspective-taking.

\subsection{Empathy Calibration in AI Systems}

Recent work has begun addressing empathy appropriateness in AI.
AI-expressed positive emotion triggers expectation-disconfirmation
rather than emotional contagion \cite{b14}. People choose human empathy
over AI empathy even when rating AI empathy as technically higher
quality \cite{b15}, and perceived empathy from AI carries less
interpersonal weight \cite{b16}. When users do not attribute mind to a
chatbot, emotional support actively undermines message effectiveness
\cite{b17}. Current LLMs show empathic abilities \cite{b2}, though
evaluation methods for computational empathy remain contested \cite{b19}.
Users' prior beliefs about an AI's motives significantly modulate
perceived empathy, trustworthiness, and effectiveness, even when the
AI's actual behavior is identical, suggesting that empathy calibration
must account for the respondent's mental model \cite{b34}.

On the computational side, the EPITOME framework identified three
empathy communication mechanisms (emotional reactions, interpretations,
and explorations) and showed that peer supporters do not self-learn
empathy over time, strengthening the case for systematic calibration
\cite{b35}. Building on this work, empathic expression can be further
differentiated by form: \emph{behavioral} (action-oriented help),
\emph{situational} (acknowledging circumstantial difficulty),
\emph{cognitive} (perspective-taking or reframing), and
\emph{experiential} (sharing one's own emotional experience) \cite{b35, b41}. These distinctions matter because different interaction
contexts may call for different empathy forms, not merely different
empathy levels. The EmpatheticDialogues benchmark noted explicitly that
``the needs for empathy have to be balanced with staying on topic or
providing information,'' the core tension this paper addresses \cite{b36}.
Empathy appropriateness has been shown to depend on context type
(instrumental vs.~experiential) and the user's functional competence
expectations, with a three-component artificial empathy framework
(perspective-taking, empathic concern, and emotional contagion)
specifying contingency propositions for when empathy creates value
versus harm \cite{b37}. A signal-cost framework further formalizes this
intuition: empathic signals are appropriate when their perceived
sincerity cost is low relative to the recipient's emotional need, but
become inappropriate when the signaling cost exceeds the supportive
benefit---as when an AI system expresses deep experiential empathy in a
context that calls for instrumental guidance \cite{b45}.

In the NLP community, the Emotional Support Conversation (ESC) framework
\cite{b18} formalized a strategy sequence (exploration, comforting,
action) for emotional support dialogues, and subsequent work has
developed systems that follow this progression \cite{b44}. However, the
ESC framework assumes the primary goal is emotional support, with
instrumental guidance as secondary. Our work addresses the complementary
case where the primary goal is decision support and empathy serves an
auxiliary function.

These findings collectively suggest that empathic expressions from AI
are processed differently from equivalent human expressions, and that
context-appropriate calibration is critical. However, this literature
has not systematically characterized the intermediate zone between
task-focused and emotional disclosure interactions.

\subsection{Support Type Matching}

Optimal matching theory proposes that support benefits depend on the
match between support type and stressor controllability \cite{b20}. In
online health communities, emotional support was most significant for
health improvement despite informational support being most prevalent
\cite{b21}. Emotional support functions as a magnifier of advice quality:
advice preceded by emotional acknowledgment is judged higher quality
\cite{b22}, a finding extended by the concept of ``advice padded with
encouragement'' \cite{b23}. These findings suggest that the appropriate
empathic register in decision-support contexts is not a simple average
of the two poles but an asymmetric profile where empathy serves as a
bridge to instrumental substance.

\section{HYPOTHESES}

Drawing on the reviewed literature, we derive five hypotheses about
empathy calibration in decision-support-under-stress interactions.

\textbf{Empathy depth.} When someone posts a purely task-focused
question (e.g., ``Should I negotiate salary before or after signing?''),
respondents typically provide direct advice with little emotional
engagement. When someone posts a purely emotional disclosure (e.g., ``I
just got laid off and feel lost''), respondents typically prioritize
emotional acknowledgment. Decision support under stress combines both
elements: the poster faces a concrete decision \emph{and} is emotionally
distressed. Because distress is present, respondents should recognize
it, producing empathy levels comparable to emotional disclosure. Yet
because a decision still needs to be made, the overall reply should not
shift entirely toward emotional engagement. In other words,
decision-support-under-stress posts should elicit empathy on par with
emotional disclosure, rather than the near-zero empathy typical of
purely task-focused exchanges.

\emph{H1: Replies to decision-support-under-stress posts will show
significantly higher empathy depth than task-focused posts, but
comparable empathy depth to emotional disclosure posts.}

\textbf{Instrumental proportion.} While H1 addresses empathy, this
hypothesis addresses the other side of the response: how much of the
reply is devoted to actionable advice and problem-solving. In
task-focused exchanges, nearly the entire reply consists of instrumental
content because no emotional need has been expressed. In emotional
disclosure, instrumentality drops because the primary need is emotional
validation, not advice. Decision-support-under-stress occupies the
middle: respondents must still deliver substantive guidance (keeping
instrumentality high), but they also allocate some response space to
emotional acknowledgment (reducing instrumentality below the
task-focused level). This predicts a three-way ordering on
instrumentality: task-focused $>$ decision support $>$ emotional disclosure.

\emph{H2: Replies to decision-support-under-stress posts will show
significantly higher instrumental proportion than emotional disclosure
posts.}

\textbf{Empathy form.} Not all empathy is expressed the same way.
Optimal matching theory \cite{b20} predicts that support type should match
the nature of the stressor: when a stressor is controllable (i.e., a
decision can be made), instrumental support is most beneficial, and
empathy should serve to facilitate that support rather than become an
end in itself. In decision-support contexts, this means empathy
functions to acknowledge the poster's difficulty and then bridge toward
advice, rather than to dwell on shared emotional experience. This
bridging function favors action-oriented forms: behavioral empathy
(offering concrete help, e.g., ``here's what worked for me'') and
situational empathy (validating the difficulty of the circumstances,
e.g., ``that's a tough position to be in''). By contrast, experiential
empathy (sharing one's own emotional reaction to a similar situation)
serves a bonding function more suited to pure emotional disclosure,
where the stressor is less controllable and felt understanding is the
primary need. We therefore expect action-oriented forms to dominate in
decision-support-under-stress replies.

\emph{H3: The dominant empathy form in decision-support-under-stress
replies will be behavioral rather than experiential.}

\textbf{Community norms.} Different online communities develop different
norms around emotional expression. A career advice community may be more
emotionally expressive than a personal finance community. If the
interaction-type effect we hypothesize is a genuine structural
phenomenon (driven by the communicative demands of the message type), it
should hold regardless of the community's overall empathy baseline. A
community with high baseline empathy should still show the same relative
ordering (decision support $\approx$ emotional disclosure >
task-focused) as a community with low baseline empathy, even if the
absolute levels differ substantially.

\emph{H4: Community norms will modulate empathy baselines while
preserving the relative ordering of interaction types within each
community.}

\textbf{Quality-empathy compensation.} Optimal matching theory \cite{b20}
further implies that when instrumental support is already strong,
additional emotional support yields diminishing returns because the
primary need (decision guidance) is already well served. Empirical work
confirms this logic: emotional support magnifies the perceived quality
of advice \cite{b22}, and ``padding'' advice with encouragement improves
reception \cite{b23}, but these effects are complementary rather than
additive. When a reply already provides high-quality instrumental
guidance (as indicated by community upvotes \cite{b24}), the bridging
function of empathy becomes less necessary because the advice speaks for
itself. When instrumental quality is lower, respondents may compensate
by investing more in emotional engagement to maintain the reply's
perceived value. This predicts an inverse relationship between reply
quality and empathic expression.

\emph{H5: Community-validated response quality will be negatively
associated with empathic expression in decision-support-under-stress
replies.}

\section{METHOD}

\subsection{Data Collection}

We collected post-reply dyads from three advice-oriented subreddits
spanning a range of empathy norms: r/careerguidance (career advice,
frequent personal/emotional context), r/Entrepreneur (business advice,
more instrumental), and r/personalfinance (financial advice, strongest
advice-giving norm). Data were collected from May 1, 2025 to July 31,
2025. Only the top-ranked reply per post was retained to capture the
community-validated ``best'' response, the reply that the community
collectively endorsed as most appropriate. This design choice privileges
the community's consensus norm over the full distribution of responses,
which is appropriate for establishing a calibration baseline. We
acknowledge that lower-ranked replies may exhibit different empathy
profiles (e.g., higher empathy from less experienced responders), and
exploring this distribution is a direction for future work.

\subsection{Post Classification}

Posts were classified into three types using automated linguistic
markers. \emph{Decision markers} captured instrumental intent through a
curated list of 23 phrase patterns (e.g., ``should I,'' ``what would you
do,'' ``torn between,'' ``deciding whether,'' ``which option'').
\emph{Emotion markers} captured emotional content via the NRC Emotion
Lexicon and affect dictionaries, including first-person emotional
predicates (e.g., ``I'm terrified,'' ``I feel stuck,'' ``I'm
overwhelmed''). To reduce false positives, emotion markers required
first-person framing (excluding third-person references) and decision
markers required interrogative or deliberative framing (excluding
declarative uses such as ``should I worry''). Classification: Type I
(task-focused) = decision markers only; Type II (emotional disclosure) =
emotion markers only; Type III (decision support under stress) = both
markers present. Posts shorter than 50 characters were excluded.

\emph{Illustrative examples} (paraphrased for anonymity):

\begin{itemize}
\tightlist
\item
  \textbf{Type I} (task-focused): ``Should I negotiate salary before or
  after signing the offer letter? Current offer is \$85K in a MCOL
  area.''
\item
  \textbf{Type II} (emotional disclosure): ``I just got laid off after
  12 years and I feel completely lost. I don't even know who I am
  outside of that job.''
\item
  \textbf{Type III} (decision support under stress): ``I'm terrified of
  making the wrong choice. I have two job offers, one is stable but
  soul-crushing, the other is exciting but risky. My family depends on
  my income and I can't sleep thinking about this.''
\end{itemize}

Type III posts are qualitatively distinct: they integrate a concrete
decision problem with emotional distress language, creating a
communicative context where respondents must address both the decision
and the affect.

\subsection{Reply Scoring}

Each reply was scored on three dimensions using two independent methods.

\emph{1) LLM-based scoring (primary):} Each dyad was scored by Claude
Sonnet (claude-sonnet-4-20250514, Anthropic) via a structured prompt
providing the full post-reply pair with explicit dimension definitions
and anchored examples for each scale point. The prompt specified:
empathy depth (0-3 ordinal: 0 = no empathic content; 1 = brief
acknowledgment such as ``that sounds tough''; 2 = moderate engagement
with the poster's emotional state; 3 = sustained empathic engagement
across multiple sentences), empathy form (multi-label: experiential,
cognitive, situational, behavioral, absent), and instrumental proportion
(0.0-1.0 continuous, defined as the share of reply content providing
actionable information, advice, or problem-solving). Structured JSON
output with retry logic ensured parseable responses. Temperature was set
to 0 across all scoring runs to minimize stochastic variation \cite{b40}.
The full scoring prompt is provided in the supplementary materials to
enable replication.

This approach follows emerging evidence that LLMs can match or exceed
human annotators on subjective text classification tasks \cite{b38, b39}, while attending to methodological recommendations for
transparency, prompt sensitivity analysis, and validation against
independent baselines \cite{b40}.

\emph{2) Pattern-based NLP (robustness check):} An independent
keyword/regex system scored the same dimensions using manually curated
lexicons (e.g., empathy phrases such as ``I understand,'' ``that must
be,'' hedging markers, and advice-giving markers such as ``you should,''
``I recommend''). This provides a model-independent baseline expected to
underestimate empathy due to inability to detect implicit empathic
signals, but critically shares no parameters, training data, or
architectural assumptions with the LLM scorer.

\emph{3) Cross-method validation:} Following recommendations for
LLM-based annotation studies \cite{b39, b40}, we treat the
convergence between these two fundamentally different scoring methods as
a triangulation strategy. Since pattern-based scoring is a known lower
bound (it cannot detect implicit empathy), we interpret agreement on
relative ordering across methods as stronger evidence than absolute
agreement on individual scores. The key validity criterion is whether
the two methods agree on the structural pattern (Type~III $>$
Type~I; Type~III $\approx$ Type~II) rather than on absolute values, a
criterion met across all hypothesis tests (Section V-G).

\subsection{Analytic Strategy}

Group differences were tested using Mann-Whitney U (pairwise) and
Kruskal-Wallis H (omnibus) tests, appropriate for ordinal/bounded
measures. Effect sizes are rank-biserial correlations. The
quality-empathy relationship (H5) was assessed via Spearman
correlations, quartile descriptives, and OLS regression controlling for
reply length and community.

\section{RESULTS}

\subsection{Corpus Overview}

The final corpus comprised 28,239 scored dyads (Table I). Type III posts
were well represented across all communities (616-2,013 per subreddit).

\begin{table}[h]
\centering
\caption{Corpus composition by community and interaction type.}
\begin{tabular}{@{}lrrrr@{}}
\toprule
Community & Type I & Type II & Type III & Total \\
\midrule
r/careerguidance & 3,911 & 2,035 & 2,013 & 7,959 \\
r/Entrepreneur & 10,302 & 2,327 & 813 & 13,442 \\
r/personalfinance & 5,312 & 910 & 616 & 6,838 \\
\textbf{Total} & \textbf{19,525} & \textbf{5,272} & \textbf{3,442} &
\textbf{28,239} \\
\end{tabular}
\end{table}

\subsection{Empathy Depth (H1)}

Table II reports the empathy depth distribution. Type III replies showed
significantly greater empathy depth than Type I ($U = 39{,}489{,}234$, $p < .001$, $r = -0.18$). Type II and Type III were not
significantly different ($U = 9{,}097{,}389$, $p = .403$). Omnibus: $H = 931.3$, $p < .001$. \textbf{H1 supported.}

\begin{table}[h]
\centering
\caption{Empathy depth by interaction type (LLM-scored).}
\begin{tabular}{@{}lrrr@{}}
\toprule
Depth & Type I & Type III & Type II \\
\midrule
0 (none) & 77.6\% & 61.1\% & 60.9\% \\
1 (brief) & 20.6\% & 31.3\% & 31.3\% \\
2 (moderate) & 1.7\% & 6.8\% & 7.0\% \\
3 (high) & 0.0\% & 0.7\% & 0.8\% \\
\textbf{M (SD)} & \textbf{0.24 (0.47)} & \textbf{0.47 (0.66)} &
\textbf{0.48 (0.66)} \\
\end{tabular}
\end{table}

\subsection{Instrumental Proportion (H2)}

Type I replies were most instrumental ($M = 0.882$), followed by Type III
($0.828$) and Type II ($0.768$). All pairwise differences were significant
($p < .001$). Omnibus: $H = 989.6$, $p < .001$. \textbf{H2
supported.}

\subsection{Empathy Form (H3)}

Table III reports form prevalence. Behavioral empathy was the most
prevalent non-absent form for Type III (36.6\%), followed by situational
(27.4\%). Experiential was lower in Type III (11.8\%) than Type II
(14.0\%). \textbf{H3 supported.}

\begin{table}[h]
\centering
\caption{Empathy form prevalence (\%, multi-label).}
\begin{tabular}{@{}lrrr@{}}
\toprule
Form & Type I & Type III & Type II \\
\midrule
Behavioral & 22.4 & 36.6 & 30.9 \\
Situational & 13.5 & 27.4 & 25.7 \\
Cognitive & 6.7 & 14.9 & 15.9 \\
Experiential & 7.0 & 11.8 & 14.0 \\
Absent & 68.6 & 48.5 & 50.9 \\
\end{tabular}
\end{table}

\subsection{Cross-Community Variation (H4)}

Table IV reports empathy depth by community and type. r/careerguidance
showed the highest empathy across all types (Type III $M = 0.57$),
r/Entrepreneur was intermediate (0.40), and r/personalfinance the lowest
(0.24). The relative ordering (Type~III $>$ Type~I; Type~III $\approx$ Type~II) was preserved within each community.
\textbf{H4 supported.}

\begin{table}[h]
\centering
\caption{Empathy depth by community and type.}
\begin{tabular}{@{}lrrr@{}}
\toprule
Community & Type I & Type II & Type III \\
\midrule
r/careerguidance & 0.386 & 0.657 & 0.572 \\
r/Entrepreneur & 0.244 & 0.424 & 0.401 \\
r/personalfinance & 0.133 & 0.207 & 0.237 \\
\end{tabular}
\end{table}

\subsection{Quality-Empathy Compensation (H5)}

Table V reports empathy by quality quartile for Type III. A monotonic
gradient was observed: Q4 (highest quality) showed lower empathy depth
(0.37 vs.~0.50 for Q1) and higher absent empathy (56.8\% vs.~46.4\%).
Spearman: $r = -.075$ ($p < .001$) for Type III, stronger than
Type II ($r = -.043$) or Type I ($r = -.020$). Median-split Mann-Whitney
confirmed lower-quality Type III replies had higher empathy depth ($M =
0.52$ vs.\ $0.41$, $p < .001$) and affective empathy ($M = 0.29$
vs.\ $0.24$, $p = .002$). \textbf{H5 supported.}

\begin{table}[h]
\centering
\caption{Empathy by quality quartile (Type III, $N = 3{,}442$).
Affect M = mean affective empathy score (proportion of reply containing
emotional reactions or experiential empathy forms).}
\begin{tabular}{@{}llrrr@{}}
\toprule
Quartile & Score range & Depth M & Affect M & Absent \% \\
\midrule
Q1 (lowest) & -14 to 1 & 0.501 & 0.283 & 46.4 \\
Q2 & 2 & 0.540 & 0.292 & 42.3 \\
Q3 & 3-6 & 0.462 & 0.285 & 49.7 \\
Q4 (highest) & 7-8,719 & 0.368 & 0.203 & 56.8 \\
\end{tabular}
\end{table}

OLS regression controlling for $\log(\text{comment length})$, type, and community
showed the coefficient on $\log(\text{comment score})$ was positive ($\beta = 0.019$,
$p < .001$) after controlling for length ($\beta = 0.216$, $p
< .001$), indicating that the compensation operates through
\emph{proportion} rather than absolute content.

\subsection{Robustness Check}

Pattern-based scoring produced convergent structural conclusions on all
five hypothesis tests: Type~III $>$ Type~I on empathy ($p < .001$); Type~II $\approx$ Type~III (ns); Type~I $>$ Type~III $>$ Type~II on instrumentality (all $p < .001$). Absolute detection differed substantially (pattern:
93-96\% at depth 0 vs.~LLM: 61-78\%), consistent with the expected
insensitivity of keyword-based methods to implicit empathic signals.
Spearman between methods: $r = .232$ ($p < .001$); exact
agreement: 73.7\%; within $\pm 1$: 97.1\%.

The moderate point-level correlation ($r = .232$) but high structural
convergence is expected given the methods' asymmetric sensitivity: the
pattern-based system detects only explicit markers, compressing most
scores to zero, while the LLM detects implicit empathy. Crucially, this
divergence is \emph{method-consistent}: the LLM systematically detects
more empathy than the pattern-based system across all interaction types,
meaning any absolute bias affects all types equally and cannot produce
the observed interaction-type differences. The structural convergence
across two methods with independent failure modes provides stronger
evidence for the relative ordering than high inter-method correlation on
absolute scores would.

\section{DISCUSSION}

\subsection{The Asymmetric Profile}

All five hypotheses were supported. Decision-support-under-stress posts
elicit an asymmetric response: empathy at the level of emotional
disclosure ($M = 0.47$ vs.\ $0.48$, ns) but instrumentality near task-focused
exchange ($0.83$ vs.\ $0.88$). This is not a simple blend; the interaction
type shapes which dimension respondents modulate. When distress
accompanies a decision problem, respondents add empathic acknowledgment
while maintaining advice-giving behavior; when distress occurs without a
decision problem, respondents shift further toward emotional engagement
and reduce instrumentality.

We note that the absolute empathy levels are low (61-78\% of replies
scored at depth 0), reflecting the advice-oriented nature of these
communities. The effect size for the Type I vs.\ Type III comparison ($r =
-0.18$) is small-to-medium by conventional standards, but this is
consistent with the fact that empathy is a secondary signal in
advice-giving contexts, exactly the point of the asymmetric profile. The
practical significance lies not in the magnitude of individual effects
but in the structural pattern: the interaction type predicts
\emph{which} dimension respondents adjust and which they hold constant,
providing a calibration template for AI systems.

\subsection{Behavioral Empathy Dominates}

Behavioral empathy (36.6\%), action-oriented help like ``have you
considered\ldots{}'' or ``here's what helped when I faced something
similar,'' was the most prevalent form in Type III. This is significant
for affective AI design: the empathic forms that human respondents
naturally use in decision-support contexts are precisely the forms that
do not require claiming emotional experience. An AI system can say
``that sounds like a difficult tradeoff'' (situational) or ``let me help
you think through the options'' (behavioral) without claiming
experiential empathy it does not possess. Convergent evidence from a
therapeutic context shows that overuse of open-ended exploratory
questions, the empathy form least prevalent in our decision-support
data, actually \emph{reduced} perceived empathy among support seekers,
while emotional validation through shared experience was most valued
\cite{b41}. This supports the view that empathy form, not just depth,
requires calibration.

\subsection{The Compensation Effect}

Higher-quality replies (by community upvotes) contain proportionally
less empathic content, particularly for Type III. This is consistent
with empathy serving a compensatory function: when instrumental value is
high, the ``bridge'' function of empathy is less necessary \cite{b22, b23}. The regression clarifies that high-quality replies are longer
and contain more of everything, but allocate a smaller \emph{share} to
emotional support. For AI design, this suggests empathic depth should
scale inversely with instrumental quality: when the system generates
strong, specific guidance, brief acknowledgment suffices.

We acknowledge that upvotes are an imperfect quality proxy: they are
subject to position bias (earlier replies receive more exposure), social
influence cascades \cite{b25}, and community-specific voting norms.
However, the OLS regression controls for reply length and community,
isolating the score-empathy relationship from these confounds. The
compensation effect is also consistent across all three communities
despite their different voting cultures, suggesting it reflects a
genuine structural pattern rather than a community-specific artifact.
The quality-empathy correlation ($r = -.075$) is modest, explaining less
than 1\% of variance, which is expected given the many other factors
that influence both upvotes and empathic content. We interpret this as a
detectable but not dominant signal, one input among many for a
calibration model, not a deterministic rule.

\subsection{Community Norms}

The 2.4x difference in Type III empathy between r/careerguidance (0.57)
and r/personalfinance (0.24) demonstrates that community norms
substantially modulate the empathy baseline. Yet the structural ordering
is preserved within every community. An empathic AI system should
calibrate not only to the interaction type but also to the community
context.

\subsection{Implications for Affective AI}

Current LLM-based systems default to uniformly high empathy across
contexts. Our findings suggest this is miscalibrated. For decision
support under stress, the appropriate register is brief situational
acknowledgment followed by structured instrumental guidance, matching
the behavioral and situational empathy forms that dominate human
responses. Inaccurate empathy is worse than minimal empathy \cite{b5},
further underscoring that calibration, not maximization, should be the
design goal. Even in fully automated CBT delivery, users valued the
chatbot's empathic ``personality'' and sense of being attended to more
than content, suggesting that a small, well-calibrated empathic signal
can be sufficient \cite{b42}. Artificial empathy creates value contingent
on context type, user expectations, and the system's functional
competence; when an AI system provides strong instrumental guidance,
excessive empathic elaboration can undermine perceived competence,
consistent with our compensation effect (H5) \cite{b37}. The signal-cost
perspective \cite{b45} offers a unifying theoretical account: in
decision-support contexts, behavioral and situational empathy carry low
signaling costs (they acknowledge difficulty without claiming shared
experience), while experiential empathy carries higher costs (it claims
emotional understanding that an AI cannot authentically possess),
explaining why human respondents---and, by extension, well-calibrated AI
systems---favor the former.

Concretely, our findings suggest a three-step operationalization for
affective AI systems: (1) \emph{classify the interaction type} of the
user's message (task-focused, emotional disclosure, or
decision-support-under-stress) using the marker-based approach validated
here; (2) \emph{select the empathy register} based on the detected type:
near-zero empathy for task-focused, high empathy for emotional
disclosure, and the asymmetric profile (brief situational/behavioral
empathy + high instrumentality) for decision support under stress; (3)
\emph{adjust for context norms}, scaling the empathy baseline to the
community or application domain (e.g., a career coaching chatbot should
operate closer to the r/careerguidance baseline, while a financial
planning tool should approximate r/personalfinance norms).

We note an important boundary condition: our baseline reflects
human-to-human empathy norms on an anonymous platform. Research on AI
empathy perception suggests that users process AI empathy differently
from human empathy \cite{b14, b15, b16, b34}, and the optimal AI
empathy level may not simply replicate the human baseline. The human
baseline provides a starting point and structural template (which forms
of empathy, which dimensions to modulate), but the absolute calibration
for AI systems requires experimental validation in human-AI interaction
contexts.

\subsection{Limitations}

Several limitations should be noted.

First, while we validate the LLM scoring instrument against an
independent pattern-based method and demonstrate structural convergence
(Section V-G), we lack a formal human inter-annotator reliability study
for the LLM-scored empathy dimensions. LLMs systematically
underrepresent emotional intensity compared to human annotators (Cohen's
$d = 0.88$) \cite{b43}, suggesting our absolute empathy scores may be
conservative. We mitigate this through our triangulation design: the key
claims rest on relative ordering across interaction types, which is
preserved across both scoring methods, rather than on absolute values.
Nonetheless, a formal validation study comparing LLM scores against
trained human coders on a subset of dyads remains an important next
step.

Second, the post classification relies on automated linguistic markers
and has not been validated against human raters. Boundary cases where
mild emotional language co-occurs with decision-seeking may be
misclassified, potentially diluting the distinctiveness of Type III. A
formal human validation study remains an important next step.

Third, community upvotes are a noisy quality proxy subject to social
influence bias \cite{b25} and position effects, though the OLS regression
controls for reply length and community. Fourth, we establish human
norms but do not test whether these norms reflect \emph{optimal}
empathy; testing optimality requires experimental manipulation in
human-AI settings. Fifth, the data are from Reddit, an anonymous,
low-investment platform; patterns may differ in identified or relational
contexts where social accountability shapes empathic expression
differently.

\section{CONCLUSION}

This paper has identified decision support under stress as a distinct
interaction type with an empirically observable empathy demand profile
positioned between task-focused exchanges and emotional disclosures.
Using 28,239 post-reply dyads from three Reddit communities, we showed
that this type elicits an asymmetric response: empathy at the level of
emotional disclosure but instrumentality near task-focused exchange.
Human respondents use behavioral and situational empathy,
action-oriented forms that serve as a bridge to advice.
Community-validated response quality is negatively associated with
empathic expression, suggesting a compensatory dynamic. These findings
establish a human baseline against which empathic AI systems can be
calibrated and suggest that context-appropriate empathy, not empathy
maximization, should be the design goal for affective computing systems.

\section*{Ethical Impact Statement}

\textit{Human subjects and data privacy.}
This research analyzes publicly available Reddit data. The study was reviewed and determined to be exempt from full IRB review under the category of research involving publicly available data with no direct interaction with human subjects. All data were collected from public posts and comments; no private messages or deleted content were included. Usernames, post IDs, and other personally identifiable information were removed prior to analysis and are not retained in the analysis dataset. Although Reddit posts are public, users may not anticipate their content being used for research. We mitigate this risk by reporting only aggregate statistics, paraphrasing all illustrative examples, and not releasing the raw dataset. The posts analyzed discuss career decisions, financial concerns, and personal stressors; while not clinical in nature, some content is sensitive, and we have handled it accordingly.

The LLM-based scoring methodology uses Claude Sonnet (Anthropic) as an annotation instrument, not as a participant. Scored post-reply texts were transmitted to the Anthropic API for annotation; no human annotators were exposed to potentially distressing content. The scoring prompt, dimensions, and validation procedures are described in Section~IV-C and the supplementary materials.

\textit{Potential negative societal impact.}
This work establishes human empathy baselines for AI calibration. We recognize that such baselines could, in principle, be misused to manipulate users through strategically deployed empathy---for example, by designing AI systems that exploit emotional vulnerability to influence decisions in commercial or political contexts. We have considered this risk carefully. The primary application of our findings is the \textit{opposite}: to reduce harm from indiscriminate empathy maximization, which existing evidence shows can trigger inauthenticity perceptions, emotional overreach, and psychological reactance \cite{b3, b4, b5, b14}. Appropriately calibrated empathy respects the user's autonomy and communicative intent rather than overwhelming it.

We also note that establishing empathy \textit{norms} (how humans actually respond) is distinct from establishing empathy \textit{prescriptions} (how AI should respond). Our baselines describe observed human behavior on an anonymous platform; they should not be adopted uncritically as design targets without further experimental validation of their effects in human-AI interaction, where the dynamics differ substantially \cite{b14, b15, b16}.

\textit{Limits of generalizability.}
Several factors constrain the generalizability of our findings. First, the data are from three English-language Reddit communities with predominantly North American users. Empathy norms are culturally variable; our baselines should not be assumed to generalize to other languages, cultures, or platforms. Second, Reddit is an anonymous, low-investment platform where social accountability is minimal; empathy expression patterns may differ in identified or relational contexts (e.g., clinical settings, workplace communication). Third, the three subreddits studied (r/careerguidance, r/Entrepreneur, r/personalfinance) represent advice-giving communities with specific demographic and topical profiles; generalization to health, relationship, or crisis support contexts requires separate investigation. Fourth, the LLM-based scoring instrument has not been validated against human coders, and LLMs have been shown to underrepresent emotional intensity relative to human annotators \cite{b43}. Our findings should be treated as context-specific baselines, not universal prescriptions for empathic AI design.

\end{document}